# Anticorrosion and biocompatibility of a functionalized layer formed on ZK60 Mg alloy via hydroxyl ion implantation

Xian Wei[a,b], Sujie Ma[a], Pinduo Liu[c], Shixiang Lu[d], Hong Qing[c], and Qing Zhao[a,e,1]

[a] Center for Quantum Technology Research and Key Laboratory of Advanced Optoelectronic Quantum Architecture and Measurements (MOE), School of Physics, Beijing Institute of Technology, Beijing 100081, China

[b] Department of science, Taiyuan Institute of Technology, Taiyuan 030008, China

[c] School of Life Science, Beijing Institute of Technology, Beijing 100081, China

[d] School of Chemistry and Chemical Engineering, Beijing Institute of Technology, Beijing 100081, China

[e] Beijing Academy of Quantum Information Sciences, Beijing 100193, China

ABSTRACT

Magnesium and its alloys have aroused tremendous interests because of their promising mechanical properties and biocompatibility. However, their excessively fast corrosion rate hinders the development of Mg alloys in the biomedical fields. Inspired by conventional ion implantation, a less-toxic functional group (hydroxyl) is used as the ion source to bombard the ZK60 Mg alloy surface to form a functionalized oxide layer. This functionalized oxide layer significantly facilitates the corrosion resistance of the ZK60 Mg alloy substrate and the proliferation of MC3T3-E1 cells, which is confirmed by electrochemical, immersion, and *in vitro* cytocompatibility tests. In comparison with results of ZK60 alloy implanted with carboxyl ions in our previous work, it is concluded that hydroxyl-treated alloys exhibit slightly higher corrosion rate while better biocompatibility. In summary, less-toxic functional ion implantation can be an effective strategy for inhibiting corrosion of Mg alloy implants and promoting their biocompatibility.

## Introduction

In addressing clinical problems related to the stress shielding and secondary operation, biodegradable materials possess significant advantages over permanent implants [1][2]. As biodegradable metals, magnesium and its alloys are widespread in orthopedic and cardiovascular applications because of their excellent biocompatibility, high strength, and elastic modulus close to that of natural bone [3]. Some Mg-based alloys have been approved by the Conformity European mark for used as bone fixation devices [4] and drug-eluting stents [5] in 2013 and 2016, respectively. An implanted bone fixation device made of Mg-5 wt.%Ca-1 wt.%Zn alloy facilitates the bone regeneration in the early stage, and then is gradually biodegraded and replaced by the newly formed bone in the long-term clinical mode [6]. In this regard, the addition of various alloying ingredients is an effective method to enhance *in vitro* and *in vivo* properties of Mg-based biomaterials. Zn is an essential element for the human body and beneficial to bone remodeling [7]. Most physiological functions are severely perturbed if there is a lack of Zn [8]. Additionally, Zr has a grain refinement effect in magnesium alloys [9] and is less toxic to cells [10]. ZK60, a type of commercial Mg-Zn-Zr series alloy, has been widely investigated because of its superior strength and ductility [11]. However, excessive Zn content may weaken the corrosion resistance of ZK60, which is a result of more second-phase particles leading to inevitable galvanic corrosion [12][13]. Inferior anticorrosion of ZK60 cannot meet *in vivo* degradation requirements, which could cause failure of mechanical support prior to full healing of the tissue [14][15].

To effectively control the corrosion process of ZK60 Mg alloy, many approaches have been developed, including alloying and surface measurements, such as ion implantation, magnetron sputtering, and coatings manufactured by polymers [16]. Alloying methods usually enhance the corrosion resistance by compressing biocompatibility [17], and coating treatments can exhibit weakened bonding force at the coating/substrate interface [18]. In contrast, ion implantation is an effective surface modification technology that directly introduces foreign ions into a substrate lattice without considering thermodynamic factors and changing the size and shape of substrates [19]. Until now, various ions have been introduced into magnesium alloy surfaces through ion implantation and have exhibited enhanced performance in anticorrosion and biocompatibility, including: (1) the metal ion systems of titanium (Ti) [20] and gadolinium (Gd) [21]; (2) the nonmetallic ion systems of nitrogen (N) [20][22] and carbon (C) [23]; and (3) the dual ion systems of Zr & N [24]

---

and Zr & O [25]. In addition to the aforementioned ion systems, organic functional ions have been presented to promote cell growth and adsorption after implantation into $Al_2O_3$ ceramics [26] and multiwalled carbon nanotubes [27]. In our previous work, we proposed the implantation of amino and carboxyl ions into AZ31 [28] and ZK60 Mg alloys [29], respectively, to study *in vitro* improvements of Mg-based materials with respect to their mechanical properties, corrosion resistance, and biocompatibility. Hydroxyl ions (another functional ion) have been exhibited excellent advantages in bioactivity and biocompatibility. Hou et al. reported enhanced bioactivity of polyhydroxy butyrate films implanted with hydroxyl ions [30]. The adhesion of nerve cells on a silicon surface was improved after hydroxyl ion implantation [31]. However, Mg-based alloys implanted with hydroxyl ions have not yet been thoroughly reported.

Herein, we fabricate a functionalized layer on the ZK60 alloy surface by hydroxyl ion implantation. The surface characterization, mechanical properties, corrosion performance, and biocompatibility of ZK60 alloy are systematically investigated before and after hydroxyl ion implantation. The performance of hydroxyl-implanted ZK60 alloy is further analyzed by comparing to the other organic ions (amino and carboxyl) implanted Mg alloys.

## 2 Materials and methods

### 2.1 Preparation of materials

ZK60 Mg alloy (Zn: 5.5 wt.%, Zr: 0.6 wt.%, Mg: balance) was cut into chips with a size of $10 \times 10 \times 2$ mm$^3$ followed by mechanical polishing using silicon carbide sandpapers ranging in grit size from 600 to 3000. The samples were ultrasonically cleaned in acetone, absolute ethanol, and distilled water for 10 min. After being dried in air, samples were bombarded with high-speed hydroxyl ions using an ion implanter (Beijing Normal University, Beijing, China) equipped with ethanol as the ion source. The implantation dosage was $0.5 \times 10^{16}$ and $1 \times 10^{16}$ ions/cm$^2$. The implanted voltage was 80 kV while keeping the pressure of the target chamber at $1 \times 10^{-3}$ Pa and the current density of the ion beam at 0.01 A/cm$^2$. To distinguish the sample treated with doses of $0.5 \times 10^{16}$ and $1 \times 10^{16}$ OH$^-$/cm$^2$, they are designated as OH-0.5 and OH-1.0, respectively.

### 2.2 Surface characterization

The chemical bonding states and depth profiles of the treated samples were identified using X-ray photoelectron spectroscopy (XPS, PHI Quantera II, Ulvac-Phi Inc.) with a sputtering rate of 25.8 nm/min based on a $SiO_2$ reference. The morphology and roughness of the samples before and after implantation were characterized using atomic force microscopy (AFM, MFP-3D-SA, Asylum Research, USA), and the corresponding data were processed using NanoScope Analysis 1.8. A nano-indenter (XP, MTS Systems Corporation, USA) was employed to analyze the surface mechanical properties (hardness and modulus) for untreated and treated samples. The hardness and modulus were recorded in real time using the vertical load scratch procedure with the indenter poking at a depth of 2 μm.

### 2.3 Electrochemical tests

An electrochemical work station (CHI760e, Chenhua Inc., China) with a classical three-electrode cell was used to evaluate corrosion properties of the samples. Specifically, a platinum electrode, saturated calomel electrode, and the tested sample were served as the counter, reference, and working electrodes, respectively. The samples were soaked in Hank's solution at 37 ℃ with an exposed area of 1.0 cm$^2$. Hank's solution, a simulated human body fluid, was composed of 8.00 g/L NaCl, 0.40 g/L KCl, 0.14 g/L CaCl$_2$, 0.10 g/L MgCl$_2 \cdot 6H_2O$, 0.10 g/L MgSO$_4 \cdot 7H_2O$, 0.06 g/L KH$_2$PO$_4$, 0.06 g/L Na$_2$HPO$_4 \cdot 2H_2O$, 0.35 g/L NaHCO$_3$, and 1.00 g/L glucose. Electrochemical impedance spectroscopy (EIS) was conducted after reaching stable open circuit potential, with a frequency ranging between 100 kHz and 100 mHz at 5 mV amplitude perturbation. Potentiodynamic polarization curves were performed at a sweep rate of 1 mV/s. The corrosion current density ($I_{corr}$) was derived from the Tafel regions of the cathodic and anodic parts of the polarization curves.

### 2.4 Immersion tests

Weight loss, pH value, and ion concentrations were monitored using immersion experiments to assess the static corrosion properties. Following ASTMG31-72 [32][32], the samples were immersed in Hank's solution with a volume area ratio of 20 mL/cm$^2$ at 37 ℃ for 1, 3, and 7 days. At each point in time, the samples were taken out of the solution and ultrasonically cleaned with chromic acid (200 g/L CrO$_3$ and 10 g/L AgNO$_3$) to eliminate corrosion products for weight loss measurements. The pH values were recorded using a pH meter at the three points in time (1, 3, and 7 days). Inductively coupled plasma optical emission spectrometry (ICP-OES, 725ES, Agilent, USA) was used to determine the concentrations of Mg, Ca, and P ions in Hank's solution. Additionally, the surface morphology of the corroded samples was examined using scanning electron microscopy (SEM, S-4800, Hitachi, Japan).

### 2.5 In vitro cytotoxicity tests

Untreated and treated samples were antecedently sterilized with ultraviolet for 2 h and then incubated in Dulbecco's Modified Eagle's Medium (DMEM, Gibco) at an extraction ratio of 1.25 cm$^2$/mL in a 5% CO$_2$ incubator at 37 ℃. After incubation for 72 h, the extracts were collected by filtering through a 0.22-μm filter and refrigerated at 4 ℃ for cell viability assay. Mouse osteoblast cells (MC3T3-E1) obtained from American Type Culture Collection were cultured in the DMEM replenished with 10% fetal bovine serum in a humidified incubator supplied with 5% $CO_2$ at 37 ℃. The cells were seeded in a 96-well plate at a density of $1 \times 10^3$ cells per well and cultured for 24 h to attachment. Subsequently, the medium was replaced by the diluted extracts with different concentrations of 100%, 70%, and 50%. The DMEM was set as the control group. MC3T3-E1 cells were cultured in the corresponding concentration of extracts for 1 and 3 days in the incubator with 5% $CO_2$ at 37 ℃. Afterwards, 10-μL Methylthiazol tetrazolium (MTT) solution was fed into each well for incubation for another 4 h. Then, 150-μL dimethyl sulfoxide was added to dissolve the formed formazan crystals. Spectrophotometric absorbance was examined using a microplate reader (Cytation3, Bio-Teh, USA) at 490 nm. The cell viability was measured using the equation: $(OD_{sample}/OD_{control}) \times 100\%$.

# 3 Results

## 3.1 Surface characterization

XPS depth profile measurement and corresponding high-resolution Mg 1s, Zn 2p, and O 1s spectra are carried out to characterize the chemical state of the treated sample with a dose of $1 \times 10^{16}$ ions/cm$^2$ (Fig. 1). As shown in Fig. 1(a), the oxygen concentration initially increases and then gradually decreases to 4.36% with increasing sputtering depth, whereas the opposite trend is observed for magnesium concentration. The higher oxygen content at the beginning of sputtering is mainly a result of surface contamination [33]. The occurrence of oxygen within a depth of 750 nm indicates that a functionalized layer rich in hydroxyl ions is formed on the treated sample surface. Moreover, the high-resolution Mg 1s spectrum in Fig. 1(b) depicts two peaks at 1302.3 and 1304.7 eV after sputtering for 5 min, which represent magnesium hydroxide [34][35] and oxide [36], respectively. The peak at 1302.3 eV shifts in the positive direction, whereas the other peak, at 1304.7 eV, migrates in the negative direction with the continuation of sputtering time. This indicates that the valence state of magnesium ($Mg^{2+}$) gradually shifts to the metallic state ($Mg^0$). The absence of Zn 2p (Fig. 1(c)) intensity at the beginning of sputtering is due to its low content near the surface after $OH^-$ ion implantation. The peaks of Zn $2p_{1/2}$ (1044.1 eV) and Zn $2p_{3/2}$ (1021 eV) are detected after sputtering for 5 min, which signifies the formation of ZnO [37][38][39]. As shown in Fig. 1(d), the peaks of O 1s are distributed mainly at 528.3 ± 0.2 eV and 530.8 ± 0.1 eV, which can be assigned to the $O^{2-}$ [40] and hydroxide/oxide (i.e., $Mg(OH)_2$, MgO, and ZnO) [41][42][43], respectively. The intensity of O 1s decreases as the sputtering continues, which is consistent with the gradual decline in O concentration. Thus, hydroxyl ion implantation generates a functionalized layer on the sample surface, which acts as a barrier to protect the ZK60 alloy from aggressive electrolyte.

It has been reported that the surface morphology of samples can be altered using ion implantation technology [44][45]. Fig. 2 depicts the surface 3D topography, derived from AFM, for samples with various implantation doses. Several hill-valley features are observed on the surface of the untreated sample. Obvious hill-valley features shift to large dome-shaped islands due to hydroxyl ion implantation. Calculated root-mean-square roughness ($R_q$) and average roughness ($R_a$) for the ZK60 Mg alloy before and after hydroxyl ion treatments are listed in Table 1. Both $R_a$ and $R_q$ are markedly reduced for the treated samples, suggesting the formation of a smoother surface due to ion implantation. This can be ascribed to reconstruction caused by the sputtering effect for surface layer atoms during the ion beam bombardment process [23][46][47]. Furthermore, treated samples with more compact and flatter surface tend to improve corrosion resistance due to the small potential difference between hills and valleys [48][49].

## 3.2 Nanomechanical property

Fig. 3 presents the change of hardness/elastic modulus, along with indentation depth, and the average hardness/elastic modulus measured in the stable range of 500-1800 nm for untreated and treated samples. The hardness/modulus of the untreated sample stays at 1.66/50.54 GPa without significant change throughout the depth. The hardness of the treated samples is comparable with that of the untreated sample within a depth of 350 nm and then becomes lower at depths exceeding 350 nm. As for the modulus, the OH-1.0 sample displays no significant difference compared with the untreated sample. The OH-0.5 sample has a decreased modulus compared with that of the OH-1.0 sample. In conjunction with the measurements of average hardness/modulus, it is evidenced that the surface mechanical properties of the samples are comparable or even slightly reduced after $OH^-$ implantation.

## 3.3 Corrosion evaluation

Fig. 4 reveals the potentiodynamic polarization results of untreated and treated samples in Hank's solution at 37 ℃. The corrosion parameters (corrosion potential ($E_{corr}$) and corrosion current density ($I_{corr}$)) determined using CorrView software are listed in Table 2. The anodic branch of the polarization curve corresponds to alloy dissolution, whereas the cathodic branch is relevant to the hydrogen evolution reaction. The treated samples possess lower $I_{corr}$ than the untreated sample. As the incident dose increased, the $I_{corr}$ gradually shifts to smaller values for treated samples. The corrosion current density, which is an important parameter for characterizing corrosion resistance of materials, is proportional to the corrosion rate. Hence, the corrosion rate is significantly reduced after hydroxyl ion implantation, which is attributed to the formation of a denser surface layer containing hydroxide/oxide.

EIS spectra are carried out to further study the corrosion behavior between the electrode and electrolyte interface. Fig. 5 depicts the Nyquist plot, Bode impedance plot, Bode phase angle plot, and equivalent circuit (EC) for all the samples in Hank's solution. As shown in Fig. 5(a), two capacitive loops are observed for the treated samples, which correspond to charge transfer at high frequency and a surface functionalized layer at low frequency, respectively. The untreated sample consists of a high-frequency capacitive loop and a low-frequency inductive loop. The occurrence of an inductive loop in frequencies below 3.61 Hz is relevant to the ion adsorption, which suggests greater vulnerability to corrosion. The diameters of the semicircle in the treated samples are larger than those in the untreated sample, which suggests that corrosion resistance of samples modified using ion implantation is greatly improved.

As shown in the Bode impedance plots in Fig. 5(b), the OH-1.0 sample achieves the highest impedance modulus at lower frequency (0.1 Hz) followed by the OH-0.5 sample, both of which have higher impedance modulus than that of the untreated sample. Furthermore, the treated samples exhibit an increased phase angle throughout the investigated frequency compared with the untreated sample. Specifically, the maximum phase angles of OH-1.0, OH-0.5, and the untreated sample are 54.2, 52.6, and 41.5, respectively. As previously reported [50][51], the impedance in the low-frequency domains represents the capacity of aggressive ions to penetrate the surface layer. Hence, higher impedance at lower frequency (0.1 Hz) suggests the formation of a more protective layer in treated samples. The enlarged phase angle implies a more capacitive response and more passive surface, which demonstrates that the current flowing through the interface is reduced and the ability of corrosive ions to penetrate through the functionalized layer is weakened after hydroxyl ion implantation. Overall, the corrosion resistance for the ZK60 Mg alloy is remarkably improved by grafting the hydroxyl ions into the sample surface, which is evidenced by the increased phase angle and impedance at low frequency.

The EC with two-time constants are used to fit experimental EIS data. As shown in Fig. 5(a), the EC model is quite different for fitting samples before and after ion implantation; the EC on the upper left is used to fit the EIS of the untreated sample, whereas that on the upper right is used for the treated samples. The electrochemical parameters extracted from the EIS spectra are summarized in Table 3. Good fits between the experimental results (drawn as scatters) and the fitted data (drawn as solid lines with corresponding color) are observed in Fig. 5(a, b) and $\chi^2$ in Table 3. $R_s$ is the solution resistance, which relies on the medium conductivity. The constant phase element, $CPE_{dl}$, represents the double layer capacitance in parallel with the charge transfer resistance ($R_t$). An inductor (L) in series with a resistance ($R_L$) is put into the EC model to interpret the inductive loop at low frequency, which may be related to local corrosion and the adsorption of corrosion products for the untreated sample [52]. As for the treated samples, $CPE_{pore}$ and $R_{pore}$ correspond to the constant phase element and film resistance associated with the constructed oxide layer using ion implantation, respectively.

The CPE component (defined by Q and n) is applied to analyze the inhomogeneity systems caused by the dispersion effect from cracks, surface oxide films, and impurities [53][54][55]. Although n = 0 or n = 1, the CPE amounts to ideal resistance or capacitance. A relationship between $CPE_{pore}$ elements (Q and n) and the thickness of the functionalized layer for treated samples can be expressed as follows [56]:

$$Q = \frac{\left(\varepsilon \varepsilon_0\right)^n}{g d \rho_d^{1-n}},$$ (1)

where $d$ is the film thickness, $\varepsilon$ is the dielectric constant, $\rho_d$ is the resistivity, and $\varepsilon_0 = 8.8542 \times 10^{-14}$ F cm$^{-1}$ is the permittivity of vacuum. Then, $d$ can be determined according to the following comparative analysis [57]:

$$\frac{d_1}{d_2} = \frac{\left(\varepsilon_1 \varepsilon_0\right)^{n_1}}{\left(\varepsilon_2 \varepsilon_0\right)^{n_2}} \frac{g_2 Q_2 \rho_{d,2}^{1-n_2}}{g_1 Q_1 \rho_{d,1}^{1-n_1}},$$ (2)

The surface properties ($\varepsilon$ and $\rho_d$) of OH-0.5 and OH-1.0 samples are considered similar because of the same incident ion species. Following Table 3, the values of $n$ are equal for the treated samples; hence, $g_1 = g_2$. Therefore, the ratio of the thickness of functionalized layers at different implantation doses can be simplified and calculated as follows:

$$\frac{d_{OH-1.0}}{d_{OH-0.5}} = \frac{Q_{OH-0.5}}{Q_{OH-1.0}} = 1.1.$$ (3)

The ZK60 Mg alloy dissolves in the electrolyte due to its activity and generates magnesium ions at the surface/electrolyte interface. This process corresponds to the electrical double layer in the EC. After ion implantation, the sample surface produces a hydroxide/oxide-rich functional layer, which effectively hinders diffusion of ions through the substrate at the interface. This process is investigated using the $CPE_{pore}$ components. As revealed in Table 3, the OH-1.0 sample exhibits a larger $R_{pore}$ value of 288 $\Omega \cdot cm^2$ than the OH-0.5 sample (259.9 $\Omega \cdot cm^2$), indicating more protective hydroxide/oxide layer for the OH-1.0 sample. This can also be supported by Eq. (3) because the functional layer on the OH-1.0 sample surface is slightly thicker than that on the OH-0.5 sample surface because of the increased implantation dose. Moreover, $R_{pore}$ is obviously larger than $R_t$ for both treated samples, which implies that the constructed functionalized layer is the primary hindrance to the penetration of electrolyte. However, as to the untreated sample, the appearance of an inductive loop at low frequency is attributed to the adsorbed species, which implies obvious local pitting corrosion. Although the value of $R_t$ (274.8 $\Omega \cdot cm^2$) for the untreated sample is larger than that for the treated samples, the anti-corrosion behavior of the treated samples can be estimated using the sum of resistance values $R_{pore}$ and $R_t$ [58]. The total values of the resistance for the OH-0.5 and OH-1.0 samples are 379.1 and 409.9 $\Omega \cdot cm^2$, respectively, which indicate that the anti-corrosion property is dramatically enhanced in the ZK60 alloy after ion implantation.

To further study the corrosion characteristics of the samples, the weight loss and pH values during immersion are shown in Fig. 6(a). After soaking for 1 day, the weight loss values are not much different between untreated and treated samples. The increase in weight loss of the untreated sample becomes significantly greater than that of the treated samples over several days, which is caused by the faster corrosion rate of the untreated alloy. Similar to the weight loss profile, the pH value for the untreated sample is higher than that for the treated samples at any immersion time. This indicates that the corrosion solution for the untreated sample contains a larger amount of hydroxide ions, forming a more alkaline environment. Fig. 6(b–d) shows Mg, Ca, and P ion concentrations of extracted corrosion solution for untreated and treated samples. The Mg ion concentration increases with the immersion time, which is caused by the dissolution of the ZK60 alloy in the corrosive solution containing chloride ions. After immersion for 7 days, the Mg ion concentration of the OH-0.5 sample increases dramatically, which may be due to the breakdown of the functionalized layer and subsequent erosion to the substrate. Nevertheless, the untreated sample still displays the highest $Mg^{2+}$ concentration at any immersion time. The Ca/P ion concentrations gradually decrease along with the immersion time, which results from the formation of insoluble corrosion products containing Ca/P. Compared with the treated samples, the untreated sample displays lower Ca/P contents, which implies the formation of more corrosion products during immersion. The increased in Mg while decreased in Ca/P ion concentrations confirm that the untreated sample possesses a faster corrosion rate in immersion tests.

SEM morphologies of sample surfaces are shown in Fig. 7. These images reveal a honeycomb-like surface at the initial stage of corrosion for both untreated and treated samples. As immersion continues, the erosion intensifies with corrosion pits spreading over the surface. After immersion for 7 days, more severe corrosion holes with large depths are visible for the untreated sample. However, only micropores and some pits appear on the surfaces of the OH-0.5 and OH-1.0 samples, respectively. Thus, a lower density of pitting and slower erosion evolution is observed for the treated samples. This reflects that the formed functional layer, by grafting hydroxyl ions onto the sample surface, provides better protection for the evolution of corrosion.

### 3.4 Cell viability

The cytotoxicity toward MC3T3-E1 cells of the untreated and treated samples is assessed in this section. Fig. 8 shows the cell viability in the extraction media concentrations of 100%, 70%, and 50% after cultured for 1 and 3 days. Both untreated and treated samples have cell viability <40% after incubation in the extract with a concentration of 100% for 1 and 3 days. Much lower viability is still observed for cells cultured with 70% and 50% extracts from the untreated sample. However, the MC3T3-E1 cells display superior biocompatibility on the treated samples, with extracted concentrations of 70% and 50%, presenting viability over 80%. The OH-1.0 sample shows the best biocompatibility with different extracts during the entire cultivation process as a result of higher implantation dose with hydroxyl ions. Overall, cell viability is significantly promoted by $OH^-$ bombardment on the sample surface, which is mainly due to the formation of a surface oxide layer that acts as a barrier to delay the dissolution of alloy ions in the extracts.

## 4 Discussion

This study successfully develops hydroxyl ion implantation into the ZK60 alloy surface and evaluates the ability of the modified layer in terms of mechanical behavior, corrosion resistance, and biocompatibility. These improvements are further analyzed by comparing other functional ion implantation Mg alloy systems regarding the aforementioned aspects.

### 4.1 Effect of ion implantation on mechanical properties

Hydroxyl ions are directly introduced into the substrate through the bombardment of high-energy ion beams, synthesizing $Mg(OH)_2/MgO$ at the near surface, within a depth of approximately 750 nm. The surface mechanical behaviors can be changed because of ion implantation. Enhanced hardness and elastic modulus are exhibited for the implantation of amino ions into AZ31 alloy [28] and carboxyl ions into ZK60 alloy [29]. However, OH-implanted samples in this study display slightly reduced hardness/modulus. This is possibly from the radiation damage effect, which gives rise to the collision process between incident ions and target atoms to deviate from the original lattice position [18][59]. This formed structural defect can spread dozens of micrometers, exceeding the implantation layer and resulting in hardness/modulus values of the treated samples maintaining lower values throughout the tested depth (Fig. 3). As shown in Fig. 3, increased values of hardness and modulus are observed for the OH-1.0 sample compared with the OH-0.5 sample. Considering the same conditions of the implantation, this difference can be attributed to the implantation dose. Overall, the hardness/modulus of Mg alloys shows little change before and after functional group (hydroxyl, carboxyl, and amino ions) implantation, indicating that the modified alloy surface maintains its original mechanical properties to some extent.

### 4.2 Effect of ion implantation on corrosion behavior

The results of both electrochemical (Fig. 4-5) and immersion tests (Fig. 6) indicate an increase in anti-corrosion for the treated samples. This improvement in corrosion resistance for the treated samples is mainly due to (1) the formation of a compound ($Mg(OH)_2$/MgO) layer, which serves as a protective layer to impede electrolyte access to the substrate; (2) the increase in solubility of the substrate by introducing species using ion implantation, which causes immiscible ions to develop a more uniform surface [19][60]; and (3) the relatively low temperature during the ion implantation process, which can be beneficial for corrosion resistance of the substrate [19].

Generally, the integrity and stability of an oxide layer can be evaluated using the Pilling-Bedworth ratio (PBR), which is the ratio of the molecular volume of the oxide to that of the metal [61]. If the PBR has a value between one and two, the oxide film on the surface is dense and protective. The oxide film is defective and inadequately covers the entire surface because of tensile stress or compressive strength when the PBR is smaller than one or greater than two [62]. Following the XPS results in Fig. 1, the surfaces of the treated samples are rich in a composite oxide layer (mainly containing $Mg(OH)_2$, MgO, and ZnO) with a depth of approximately 750 nm. In contrast, a limited amount of magnesium oxide layer is generated by natural oxidation for the untreated ZK60 alloy surface. As indicated by the PBR values listed in Table 4 [63][64], $Mg(OH)_2$ and ZnO have greater protection with higher surface coverage than MgO. This indicates that the treated samples hinder corrosion from the electrolyte to a degree as compared with the untreated sample. Initially immersed in Hank's solution, Mg dissolves to produce electrons at the anodic electrode, accompanied with $H^-$ consuming electrons, to release $H_2$ and pH values (Fig. 6(a, b)). Then, the released $Mg^{2+}$ reacts with $OH^-$, leading to the formation of $Mg(OH)_2$. This low-solubility corrosion product, along with the hydroxide/oxide-containing layer, enhances resistance of the treated samples to penetration of negative ions, such as $Cl^-$. However, the untreated sample without a functionalized layer has difficulty resisting the invasion of electrolyte, leading to serious corrosion holes (as reflected in Fig. 7). The barrier properties of the hydroxide/oxide-containing layers of the treated samples indeed reduce the corrosion rates compared with those of the untreated sample, although $Mg(OH)_2$ tends to decompose because of its instability at a pH value of 11.5 in Hank's solution [65][66].

### 4.3 Effect of ion implantation on biological performance

Both Mg and Zn are essential elements for the human body, but the daily allowable dosages of the two elements vary greatly. Mg is biocompatible, with a much higher allowable dosage of 320-400 mg/day [67], whereas Zn has an allowable dosage of 12-15 mg/day [68][69]. Consequently, $Zn^{2+}$ ions released in body fluid facilitate activity, proliferation, and adhesion of cells at a lower concentration, whereas higher concentrations of $Zn^{2+}$ do the opposite [1]. As shown in Fig. 8, the 100% extracts of all samples exhibit toxic effects on MC3T3-E1 cells, mainly resulting from the higher amount of Zn in the extracts. Jin et al. [70] and Hong et al. [71] reported similar *in vitro* cytotoxic effects on MC3T3-E1 cells from ZK60 and ZK40 alloys. Nevertheless, the behavior of 100% extracts does not thoroughly reflect *in vivo* biocompatibility. This is because (1) the ion concentration of the *in vivo* implants decreases, because of ion diffusion in the circulatory system, and (2) the corrosion rate of implants *in vivo* is much lower than that of *in vitro* simulations. Therefore, the diluted extracts of the treated alloys possess no cytotoxicity effects, which is comparable with *in vivo* behavior [65]. The treated samples show viable cells compared with the untreated sample. The reason behind these results is relevant to the corrosion rate. The corrosion resistance of the treated samples is enhanced (as shown in Fig. 6), leading to reduced ion concentration of Mg and Zn released in the medium, which facilitates cell growth. However, the cytotoxicity is still visible in the 50% diluted extract for the untreated sample. This implies that too high a concentration of metal ions, such as $Zn^{2+}$, is released in the medium, which affects cell performance. Moreover, severe localized corrosion (as displayed by the SEM images in Fig. 7) will influence the growth and adsorption of cells [7] and stimulate the formation of a thick layer of fibers [1] if the untreated sample is directly implanted into the human body.

### 4.4 Comparison with other functional group ion implanted alloys

*In vitro* weight losses of ZK60 alloys implanted with hydroxyl and carboxyl ions calculated from the immersion test are presented in Fig. 9(a) for comparison. The weight losses of all hydroxyl-treated samples are higher than those of the carboxyl-treated samples after immersion for 1 day in Hank's solution. The OH-1.0 sample possesses a reduced value of weight loss compared with the CO-1.0 sample for 3 and 7 days, suggesting more superior corrosion resistance of the hydroxyl-treated sample with the same dose of $1.0 \times 10^{16}$ ions/cm$^2$. The corrosion rate of the CO-5.0 sample is the lowest, reflected by minimal weight loss in the immersion experiment, which indicates that a significant increase in the implantation dose may possibly improve corrosion resistance.

Fig. 9(b) shows an overall comparison of biocompatibility among the Mg alloys implanted with various organic functional ions. Compared to the AZ31 alloys, both untreated and hydroxyl/carboxyl treated ZK60 alloys exhibit cell viability less than 60% for the extract with concentration of 100%. This difference may be related to the relatively high Zn content in ZK60, leading to visible cytotoxicity effect [13][72]. Compared with the carboxyl-treated ZK60 alloys, hydroxyl-treated samples have more viable cells, especially the OH-1.0 sample. Overall, organic functional group ion implantation improves corrosion resistance and biocompatibility of magnesium alloy substrates. Moreover, the increase in corrosion resistance of hydroxy treated samples is not as good as that of carboxyl-treated samples, but the biological performances are better than those of carboxyl-treated samples.

## 5 Conclusions

In this study, the hydroxyl functional group is implanted into ZK60 alloy surfaces to form a relatively smooth layer rich in oxide (Mg(OH)$_2$/MgO). The feasibility of this implanted material to be utilized as biodegradable bone material is evaluated through investigation of surface mechanical properties, corrosion performance, and biocompatibility. Both surface hardness and elastic modulus gradually decrease after ion implantation due to damage effects. Hydroxyl-treated samples display reduced corrosion rate, as reflected using electrochemical techniques and immersion tests, compared with the untreated sample due to the formation of a passive barrier on the implanted layer. The surface-modified layer significantly promotes biocompatibility of ZK60 Mg alloys. In comparison with carboxyl-treated ZK60 alloys in literature [29], the hydroxyl-treated samples achieve slightly lower corrosion resistance but higher biocompatibility. Our investigations are of great significance for understanding the corrosive and biological behaviors of Mg alloys implanted with hydroxyl ions in a simulated human environment. This study provides a new perspective and research basis on directly grafting various types of ions (such as organic functional groups) into Mg implants.

## Conflict of Interest Form

The authors declare no conflicts of interest.

## Acknowledgments

This work is supported by the National Science Foundation (NSF) of China, grant No. 11675014. Additional support was provided by the Ministry of Science and Technology of China (2013Y Q030595-3).

Table 1 Roughness (nm) derived from the AFM analysis of untreated and treated samples.

| Sample | $R_a$ (nm) | $R_q$ (nm) |
|---|---|---|
| Untreated | 42.7 | 52.8 |
| OH-0.5 | 33.1 | 41.2 |
| OH-1.0 | 26.3 | 34.3 |

Table 2 Corrosion potential ($E_{corr}$) and current density ($I_{corr}$) derived from the polarization curves.

| Sample | Untreated | OH-0.5 | OH-1.0 |
|---|---|---|---|
| $E_{corr}$ (V vs. SCE) | −1.585 | −1.607 | −1.600 |
| $I_{corr}$ ($\mu A/cm^2$) | 291.94 | 103.93 | 88.01 |

Table 3 Fitted EIS results based on the EC models.

| Sample | Untreated | OH-0.5 | OH-1.0 |
|---|---|---|---|
| $R_s$ ($\Omega \cdot cm^2$) | 30.02 | 21.27 | 19.73 |
| $Q_{pore}$ ($\Omega^{-1}\ cm^{-2}\ S^n$) | / | $2.6 \times 10^{-5}$ | $2.38 \times 10^{-5}$ |
| $n_{pore}$ | / | 0.9 | 0.9 |
| $R_{pore}$ ($\Omega \cdot cm^2$) | / | 259.9 | 288 |
| $Q_{dl}$ ($\Omega^{-1}\ cm^{-2}\ S^n$) | $7.32 \times 10^{-5}$ | $1.35 \times 10^{-4}$ | $1.25 \times 10^{-4}$ |
| $n_t$ | 0.79 | 0.78 | 0.78 |
| $R_t$ ($\Omega \cdot cm^2$) | 274.8 | 119.2 | 121.9 |
| L (H cm$^{-2}$) | 875.6 | / | / |
| $R_L$ ($\Omega \cdot cm^2$) | 517 | / | / |
| $\chi^2$ ($\times 10^{-3}$) | 3.6 | 3.3 | 3.8 |

Table 4 Pilling-Bedworth ratio (PBR) values of the involved oxides.

| Compound | MgO | $Mg(OH)_2$ | $MgCO_3$ | ZnO |
|---|---|---|---|---|
| PBR | 0.8 | 1.8 | 2.04 | 1.57 |

Fig. 1 (a) XPS depth profile and (b-d) high-resolution spectra at different sputtering times for the treated sample with a dose of $1 \times 10^{16}$ OH$^-$/cm$^2$.

Fig. 2 AFM images of the various sample surfaces: (a) untreated, (b) OH-0.5, and (c) OH-1.0.

Fig. 3 (a) Hardness, (b) elastic modulus curves, and (insert) average values measured from 500 to 1800 nm, of the untreated and treated samples.

Fig. 4 Potentiodynamic polarization plot of the untreated and treated samples in Hank's solution.

Fig. 5 (a) Nyquist plots with equivalent circuits inserted and (b) Bode plots of EIS spectra for the untreated and treated samples. The fitted data are plotted as solid curves with corresponding colors.

Fig. 6 (a) Weight loss and pH values for the untreated and treated samples after immersion; (b–d) concentrations of Mg, Ca, and P ions released into the Hank's solution.

Fig. 7 Surface morphologies of untreated and treated samples during immersion for 1, 3, and 7 days.

Fig. 8 MC3T3-E1 cell viability with different concentrations of alloy extracts.

Fig. 9 Comparison of (a) weight loss and (b) cell viability among various functional groups implanted with Mg alloys: CO-1.0 and CO-5.0 represent ZK60 Mg alloy implanted with COOH$^+$ at doses of $1 \times 10^{16}$ and $5 \times 10^{16}$ ions/cm$^2$, respectively; NH-1.0, NH-5.0, and NH-10 represent AZ31 Mg alloy implanted with NH$_2^+$ at doses of $1 \times 10^{16}$, $5 \times 10^{16}$, and $10 \times 10^{16}$ ions/cm$^2$, respectively.

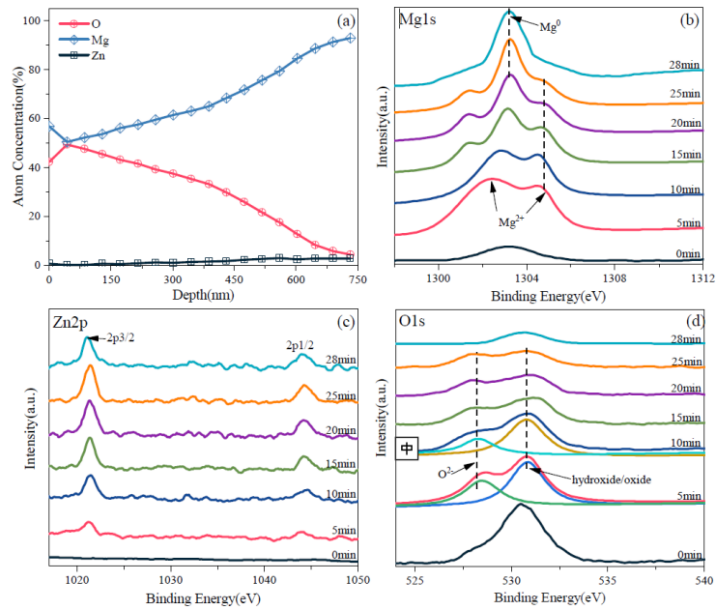

Fig. 1 (a) XPS depth profile and (b-d) high-resolution spectra at different sputtering times for the treated sample with a dose of $1 \times 10^{16}$ OH$^-$/cm$^2$.

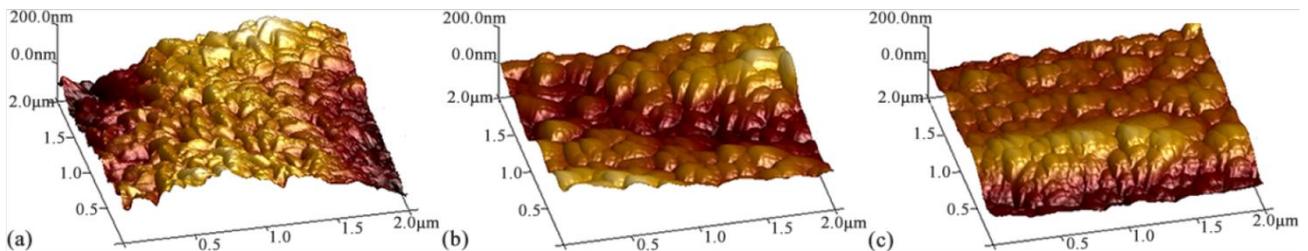

Fig. 2 AFM images of the various sample surfaces: (a) untreated, (b) OH-0.5, and (c) OH-1.0.

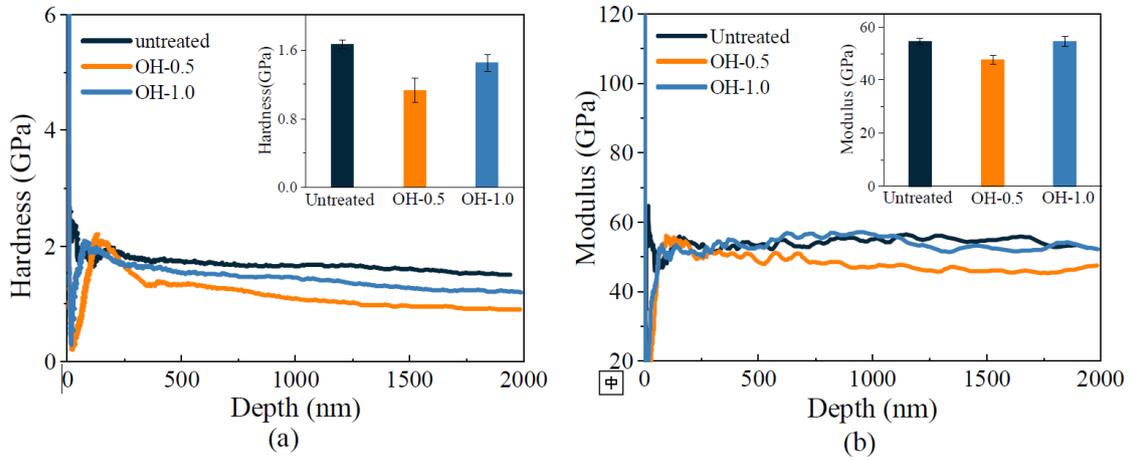

Fig. 3 (a) Hardness, (b) elastic modulus curves, and (insert) average values measured from 500 to 1800 nm, of the untreated and treated samples.

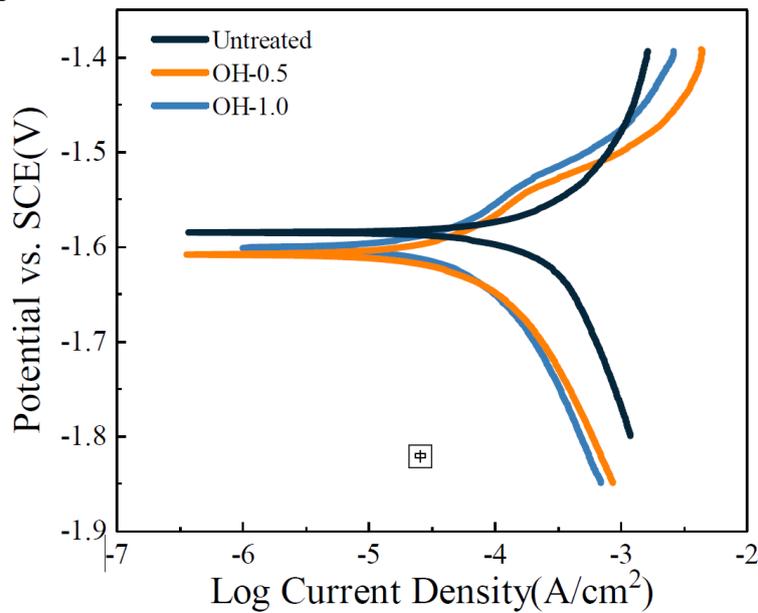

Fig. 4 Potentiodynamic polarization plot of the untreated and treated samples in Hank's solution.

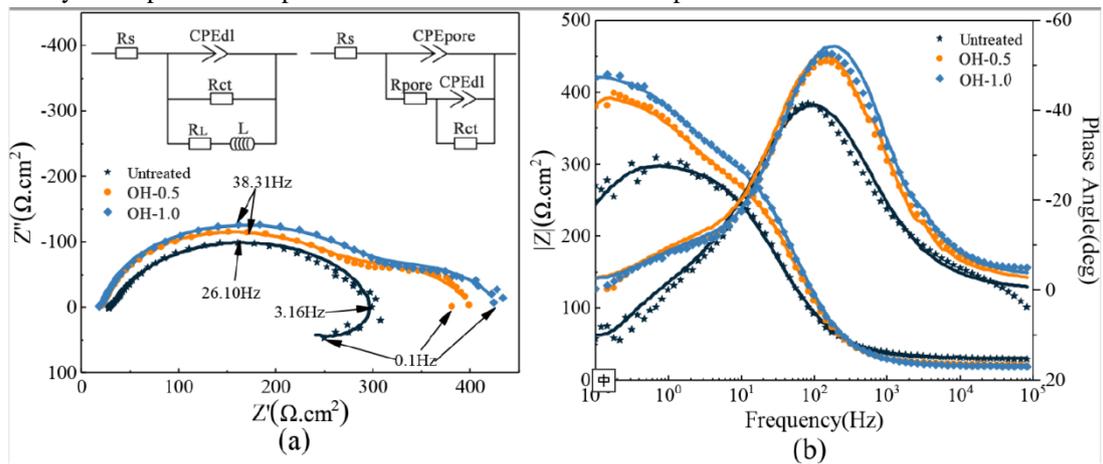

Fig. 5 (a) Nyquist plots with equivalent circuits inserted and (b) Bode plots of EIS spectra for the untreated and treated samples. The fitted data are plotted as solid curves with corresponding colors.

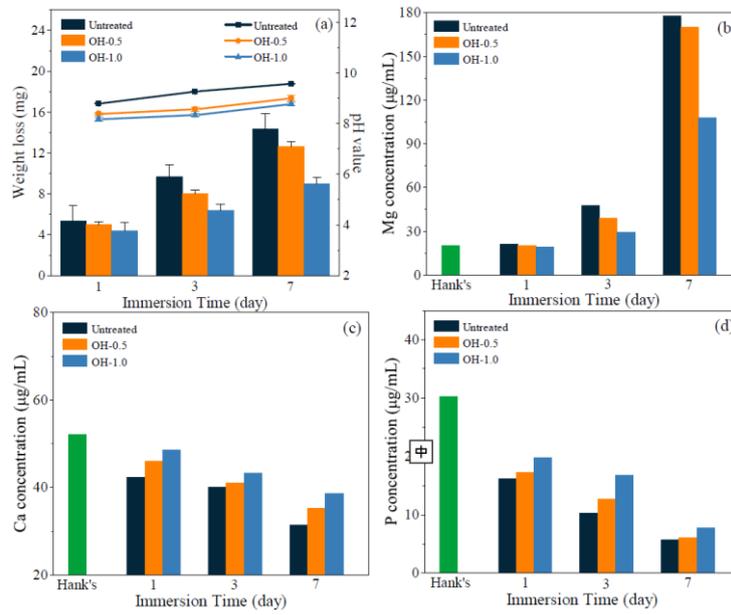

Fig. 6 (a) Weight loss and pH values for the untreated and treated samples after immersion; (b–d) concentrations of Mg, Ca, and P ions released into the Hank's solution

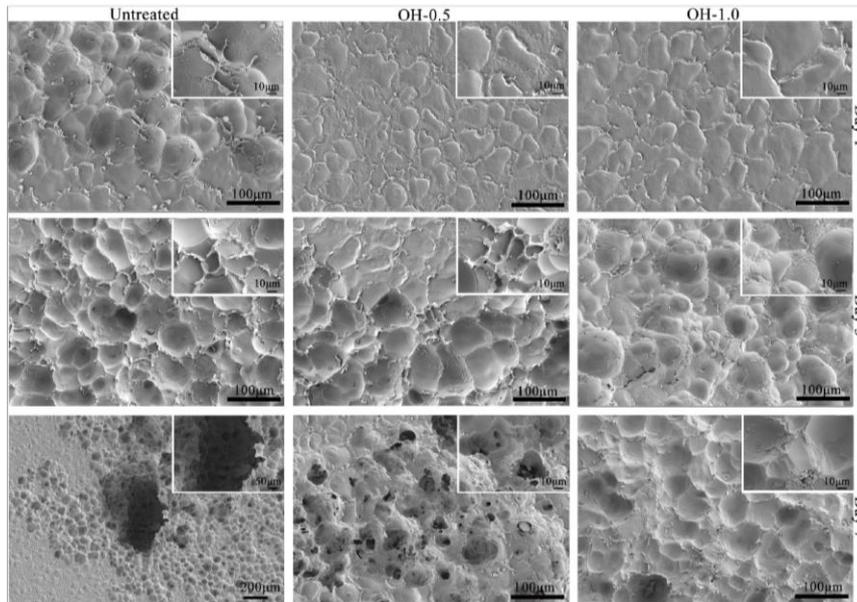

Fig. 7 Surface morphologies of untreated and treated samples during immersion for 1, 3, and 7 days.

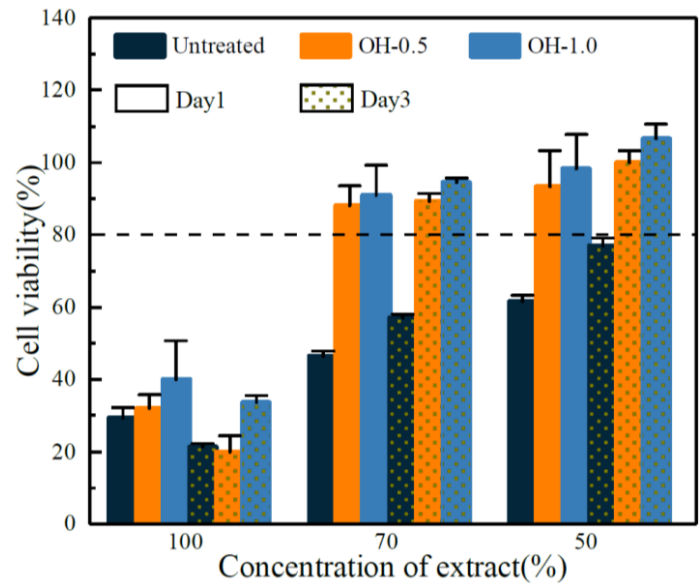

Fig. 8 MC3T3-E1 cell viability with different concentrations of alloy extracts.

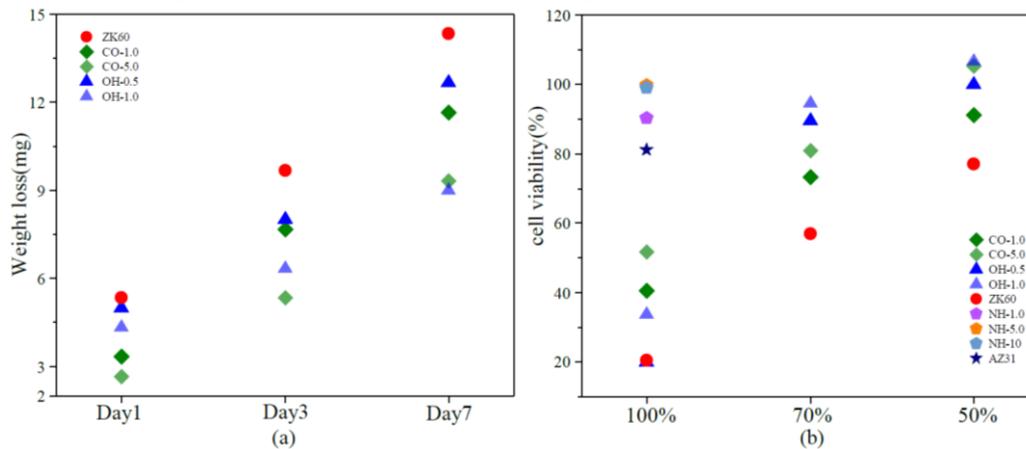

Fig. 9 Comparison of (a) weight loss and (b) cell viability among various functional groups implanted with Mg alloys: CO-1.0 and CO-5.0 represent ZK60 Mg alloy implanted with $COOH^+$ at doses of $1 \times 10^{16}$ and $5 \times 10^{16}$ ions/cm$^2$, respectively; NH-1.0, NH-5.0, and NH-10 represent AZ31 Mg alloy implanted with $NH_2^+$ at doses of $1 \times 10^{16}$, $5 \times 10^{16}$, and $10 \times 10^{16}$ ions/cm$^2$, respectively.